\documentclass[twocolumn,showpacs, preprintnumbers]{revtex4}
\usepackage{amssymb}
\usepackage{graphicx}
\usepackage{dcolumn}
\usepackage{bm}
\usepackage{cancel}
\usepackage{footmisc}

\input{tcilatex}
\begin{document}

\title{Catastrophic shifts and lethal thresholds in a propagating front \\
model of unstable tumor progression}
\author{Daniel R. Amor$^{1,2}$, Ricard V. Sol\'e$^{1,2,3}$}
\affiliation{$^1$ ICREA-Complex Systems Lab, Universitat Pompeu Fabra, Dr. Aiguader 80,
08003 Barcelona, Spain\\
$^2$ Institut de Biologia Evolutiva, CSIC-UPF, Psg Barceloneta, Barcelona,
Spain\\
$^4$ Santa Fe Institute, 1399 Hyde Park Road, New Mexico 87501, USA}

\begin{abstract}
Unstable dynamics characterizes the evolution of most solid tumors. Because
of an increased failure of maintaining genome integrity, a cumulative
increase in the levels of gene mutation and loss is observed. Previous work
suggests that instability thresholds to cancer progression exist, defining
phase transition phenomena separating tumor-winning scenarios from tumor
extinction or coexistence phases. Here we present an integral equation
approach to the quasispecies dynamics of unstable cancer. The model exhibits
two main phases, characterized by either the success or failure of cancer
tissue. Moreover, the model predicts that tumor failure can be due to either
a reduced selective advantage over healthy cells or excessive instability.
We also derive an approximate, analytical solution that predicts the front
speed of aggressive tumor populations on the instability space.
\end{abstract}

\maketitle



\section{Introduction}

Cancer is a disease that can be initiated by the failure of a single cell.
Whenever such failure leads to some proliferation advantage over the
neighboring somatic cells, this single cell is prone to originate its own
cell lineage within its host tissue. After many rounds of replication,
additional failures may occur, eventually generating a large population of
abnormal, proliferating cells. This would be a rough description of the
disease, but it would be more appropriate to say that cancer is an
evolutionary dynamic process [1,2]. Changes occur in time and accumulate
over generations and the final success of the tumor requires an appropriate
accumulation of changes affecting different types of genes.

We can classify cancer genes into three basic categories [3]: (a) oncogenes,
(b) tumor suppressor genes and (c) stability-related genes. These groups
corresponds to genes that (a) increase replication due to mutation, (b)
increase cell growth when the gene is silenced or lost and (c) modify genome
stability due to failures in cell division, repair and maintenance
mechanisms [4-9]. All these changes occur through the process of cell
replication, when cancer genes are likely to experience mutations or losses
[10] leading to the emergence of fitter mutant clones.

In order to understand the evolution of cancer, a huge amount of
mathematical models have analyzed the impact of selection [11] on the
evolution of clones. The stochasticity of mutations has also been shown to
play a major role in triggering the clonal competition among different
mutants, and could be a principal reason for the high heterogeneities (and
the long waiting time to malignancy) observed in cancer development [12].
Another very relevant mechanism is spatial structure, which is also very
significant in many ecological and evolutionary processes [13]. For example,
in the context of asexual evolving populations, the spatial competition
between different clones slows down the establishment of driver mutations
(i. e. mutations causing a selective advantage) [14] if the population
exceeds a critical size. Analogous results have been found in the context of
cancer evolution, where space can increase the waiting time to tumor
malignancy [15]. In other cases, the spatial invasion of tumors has been
modeled as a propagating front [16]. This has permitted to, e.g., compare
the role of advection (chemotaxis) and cell diffusion on the invasion speed
of glioblastomas [17,18], or analyze the invasiveness enhancement by acidic
pH gradients at tumor-host interfaces [19].

Although most classic models of cancer evolution deal with those factors
associated with growth and competition among clones, a specially important
characteristic of most tumors is precisely the increased levels of
instability associated to progression. Instability can be understood in
terms of mutations but also of losses and gains of genetic components that
modify genome stability, making cells more prone to errors while replicating
[20]. Mutations have been an intrinsic part of all evolutionary models of
population dynamics (including cancer) but it is typically assumed that
mutation rate remains constant over time. In genomically unstable tumors,
the failure of the repair mechanisms, along with the generation of
aneuploidy, makes possible to damage key components associated to the
maintenance of genome integrity [4-9,20].

With their loss or failure, further increases of instability are expected to
occur, since other genes linked to stability and repair are more likely to
be damaged. As a consequence, instability itself can evolve over time. Such
evolvable trait raises the question of how much instability can accumulate
through carcinogenesis. It has been suggested that optimal instability rates
[4] as well as thresholds to instability exist. The latter define the
transition boundaries between viable and non-viable cancer populations
[21-24]. There are actually examples of phase transitions defining the
boundaries of viability in RNA viruses[25-30]. RNA virus populations are
quasispecies [22,30] i.e. highly heterogeneous, related genotypes. Critical
thresholds of mutation have been predicted and later experimentally tested
[31-33] using in-vitro scenarios. The presence of such critical transitions
have also received a great interest from the field of statistical physics.
The nature of the resulting phase transitions have been analyzed for several
fitness landscapes, both for finite-size competing molecules [34] and in the
limit of infinitely large chains [35]. Error thresholds have also been
reported in asexual evolutionary scenarios beyond the RNA viruses.
Remarkably, it has been reported that natural selection, favoring immediate
fitness benefits, may permit the hitchhiking of deleterious mutations that
will finally lead to the population's extinction in the long term[36].

The similarities between unstable cancer and RNA viruses suggests a
therapeutically very interesting possibility: the use of additional
instability as anticancer therapy [23,37]. That means that, instead of
trying to decrease the tumor cells' mutagenesis, an attempt to increase it
towards non-viable levels could be a suitable way to fight the disease. Due
to the qualitatively sharp change associated to the presence of instability
thresholds, a physics approach to phase transitions in cancer quasispecies
can be successfully used [23,24,38-40]. In this paper we explore the
dynamics and phases of unstable cancer by constructing an analytical model
of tumor progression to be defined as a front propagation problem [41] in
the space of instability. By using this approximation we provide a better
and easily extendable formal description of tumors that allows to
characterize both the presence of transitions and the population structure
that emerges in each phase. It also provides a well defined, formal approach
to predict the speed of cancer propagation.

The paper is organized as follows. In section II we present the rationale
for the presence of a phase transition phenomenon separating a phase where
the tumor will fail to succeed due to a high instability from another phase
where it is expected to win. In section III we revisit the previous linear,
discrete model of cancer cells dynamics and we explain some of its
limitations. Section IV is devoted to present the integral model of unstable
cancer, that improves the previous mathematical description of the disease.
Section V presents several scenarios for tumor evolution predicted by the
integral model (an analysis of the resulting phase space is included). In
section VI we derive an approximate, analytical expression for the tumor
front speed on the instability space, and we compare it with some numerical
solutions for the model equations. The last section is devoted to discuss
the potential implications of our results.

\section{Transitions in tumor instability}

In order to provide a rationale for the existence (and potential
implications) of instability thresholds, let us first consider a mean field,
two-compartment model of unstable cancer dynamics. In this model the
population will be composed of two cell species, namely, host cells $H$ and
cancer cells $C$. If we indicate as $r_{n}$ and $r_{c}$ the rates of growth
of normal (host) and cancer cells, respectively, we can write the following
evolution equations: 
\begin{eqnarray}
{\frac{dH}{dt}} &=&r_{n}H-H\phi (H,C) \\
{\frac{dC}{dt}} &=&r_{c}C-C\phi (H,C),
\end{eqnarray}
where $\phi (H,C)$ is an outflow term that represents the competition
between both species. If we consider that the overall cell population $H+C$
is constant (because cells fill a given fixed space) the function $\phi$
reads $\phi =r_{n}H+r_{c}C$ which is actually the average rate of growth.

The following step consists in defining the growth rates for each species.
Regarding normal cells, it is sensible to assume a constant growth rate $%
r_{n}$ (normal cells are renewed in a stable way to ensure that body
functions are properly carried on). The situation is different for cancer
cells, for which their average growth rate $r_{c}$ will depend on how much
mutations have been accumulated in the cells' genome. Concretely, the growth
rate $r_{c}$ can be increased by the effects of driver mutations (i.e.,
mutations promoting cell replication) or decreased by deleterious mutations.
Let us denote $\mu $ as the probability that a mutation takes place when
replicating a given gene. Thus, $\mu $ is a measure of the genetic
instability of the population. Consider that there exist a number $N_{r}$ of
growth-related genes. If a growth-related gene is damaged (mutated) during
the cell replication process, an average increase $\delta _{r}$ in the
growth rate is expected. Thus, the average increase in $r_{c}$ due to driver
mutations affecting our cancer population is simply: 
\begin{equation}
f_{1}(\mu )=N_{r}\mu \delta_{r}.
\end{equation}
Similarly, we should expect a decrease in the growth rate due to the
potential damage produced if a house keeping gene is damaged or lost. If $%
N_{h}$ indicates the number of such genes, the probability that no one is
damaged will read 
\begin{equation}
f_{2}(\mu )=(1-\mu )^{N_{h}}.
\end{equation}
Available estimates indicate that $N_h \sim 500-600$ essential genes exist
[42] whereas $N_r$ can be smaller or higher depending on the type of cancer
considered. Around one percent of genes in the human genome appear related
to the emergence of cancer [43].

In the above eqs. (3) and (4) we have assumed that the mutation and
replication rates are the same for all genes, as well as a constant fitness
benefit $\delta _{r}$ from driver mutations. Obviously, this is only simple
approach to the more complex reality in which each gene mutates with a
different probability and provides a different fitness benefit (or loss).
Thus, the final rate of replication will be the product: 
\begin{equation}
r_{c}(\mu )=f_{1}(\mu )f_{2}(\mu )=\left( r_{n}+\mu N_{r}\delta r\right)
(1-\mu )^{N_{h}},
\end{equation}
where we have assumed that, in the absence of genetic instability ($\mu =0$)
the normal cells replication rate $r_{n}$ is recovered. The above function
(5) has a maximum at a given optimal instability rate. This is shown in Fig.
1, where we plot $r_{c}(\mu )$ for a given combination of parameters. The
maximum is achieved at an optimal instability level $\mu _{o}\approx 1/N_{h}$%
.

\begin{figure}[tbh]
\begin{center}
\includegraphics[width=0.45 \textwidth]{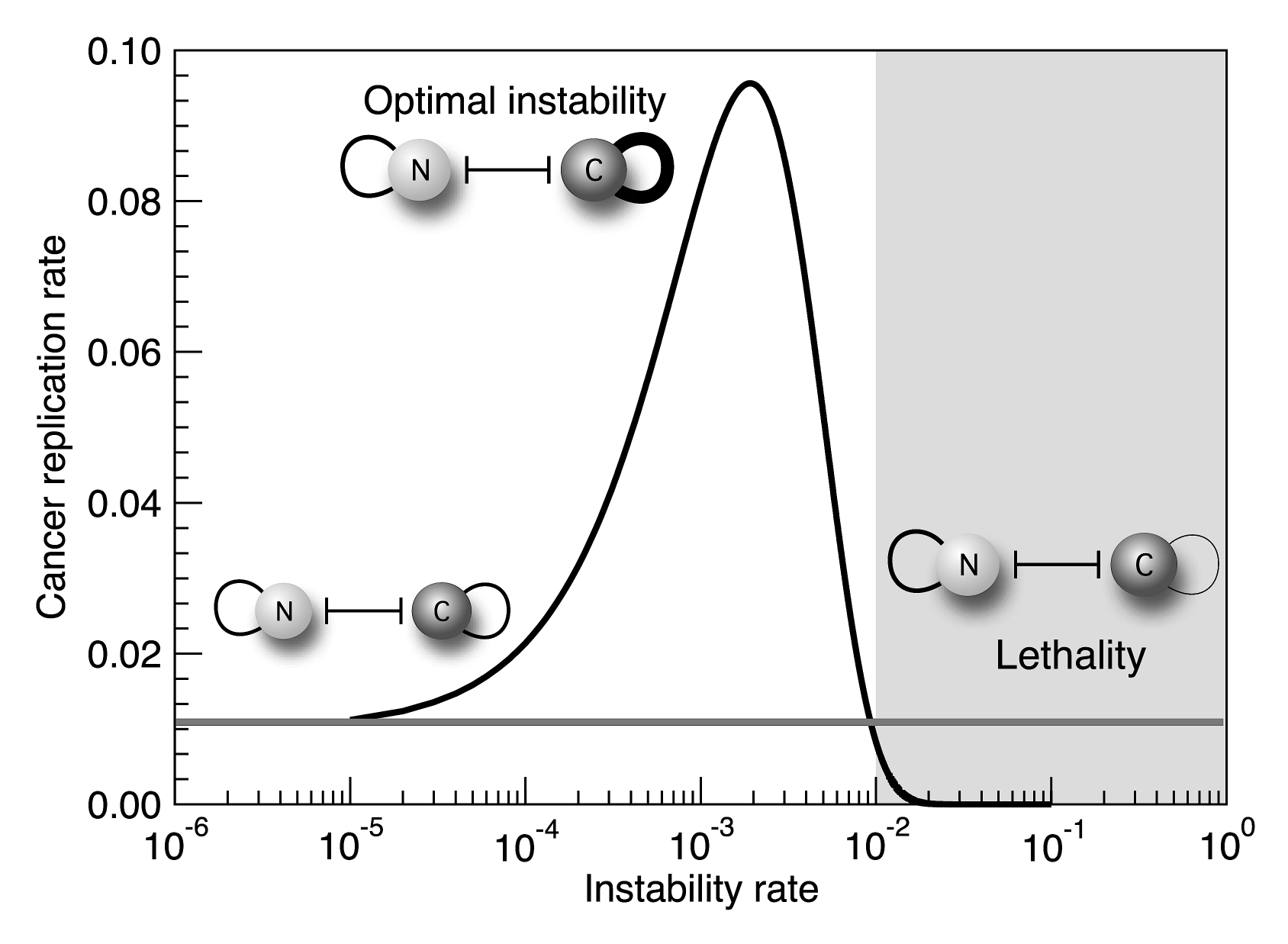}
\end{center}
\caption{Optimality and lethality in unstable cell populations. The vertical
axis indicates the cancer replication rate against instability, as predicted
from equation (5). The replication rate of normal cells is $r_{0}=0.01$ (in
arbitrary units). The cancer population is assumed to be homogeneous. At low
instability rates, competition between the two cell populations is symmetric
and cancer coexists or slowly grows. The peak at the optimal rate $\protect%
\mu_o$ is associated to the fastest potential growth of cancer. The grey
area indicates the lethal phase, where excessive instability leads to a
reduced proliferation of cancer cells, which are overcompeted by normal
cells.}
\label{fig1}
\end{figure}

Considering Eq. (5) and the constant population constraint, it is possible
to reduce the system of Eqs. (1)-(2) into a single equation describing the
dynamics of the system in terms of the genetic instability $\mu $. Thus, the
cancer cell population is now captured by a logistic-like nonlinear
equation: 
\begin{equation}
{\frac{dC}{dt}}=r_{n}(\Gamma (\mu )-1)C(1-C)
\end{equation}
Two fixed points are present: the zero-population one $C^{\ast }=0$ and the
maximum population state, here $C^{\ast }=1$. It is easy to see that the
first is stable if $\Gamma (\mu )<1$ and unstable otherwise. By properly
defining the function $\Gamma (\mu )$ we might be able to define the
conditions under which genetic instability allows cancer growth to occur and
overcome the host tissue. The critical mutation rate separating the two
scenarios is sharp and defines a phase transition.

The presence of a phase transition in this toy mean field model involving
competition between two homogeneous populations offers an interesting
prediction: further increases of instability can force cancer cells to enter
the lethal phase. However, understanding how such shifts can occur requires
a better understanding of the ways cancer cell populations evolve. Cancer
cell populations are highly heterogeneous [44,45] and that means that we
need to depart from the previous model approach.

\begin{figure}[tbp]
\begin{center}
\includegraphics[width=0.45 \textwidth]{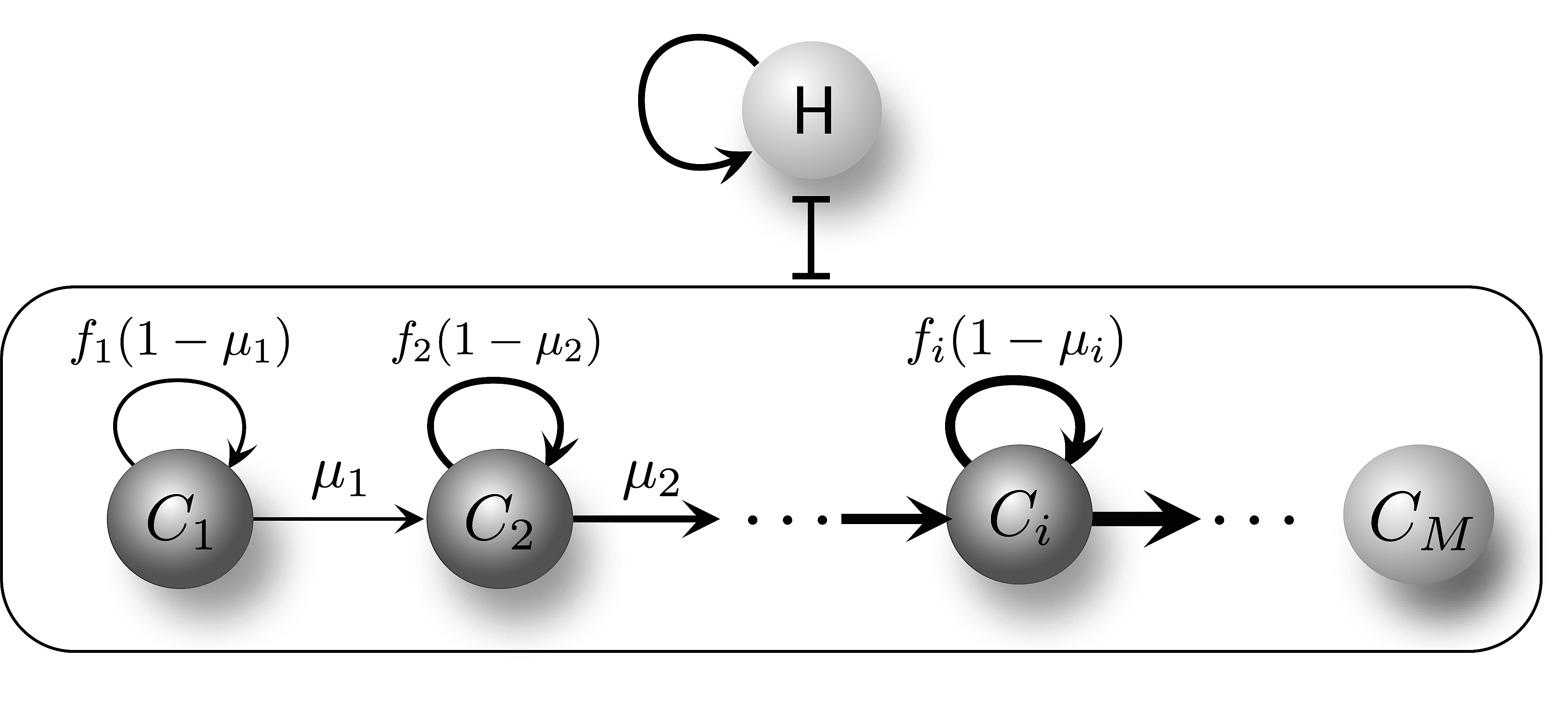}
\end{center}
\caption{Linear model of competition between normal cells ($H$) and a
heterogeneous population of cancer cells, indicated as $%
C_{1},C_{2},...,C_{i} $ which replicate with increasing rates $f_i$ and
mutate also at faster rates $\protect\mu_i$, as highlighted by the
increasingly thick arrows. The effective replication rate of a given $C_k$
compartment is $f_k(1-\protect\mu_k)$.}
\label{fig2}
\end{figure}

\section{Linear model of unstable cancer}

In an early paper [46] a discrete, sequential model of unstable cancer was
introduced. The model considered a population of cancer cells having
different levels of instability and competing among them and with the normal
tissue (figure 2). This led to a description of $M$ levels of instability describing an heterogeneous cancer cells population,
which was governed by the following set of $M$ differential
equations:%
\begin{equation}
\frac{dC_{i}}{dt}=f_{i-1}\mu _{i-1}C_{i-1}+f_{i}(1-\mu _{i})C_{i}-C_{i}\Phi
(H,\mathbf{C}),
\label{2}
\end{equation}
where $i=1,3,...,M$, $\mathbf{C}=(C_{1},...,C_{M})$ and we consider the 
terms $\mu_{0}=\mu_{M}=0$ so that they properly define the first 
and last of the equations in (7). In the set (7), $H$ indicates the host (healthy) population, whose dynamics would be
described by an additional equation $dH/dt=f_{H}(H,C)$ which takes the
general form 
\begin{equation}
\frac{dH}{dt}=G(H)-H\Phi (H,\mathbf{C})  \label{10}
\end{equation}%
Here $G(H)$ introduces the explicit form of growth characterizing the normal
tissue. A constant population constraint (CPC) was also introduced, namely a
total constant population size $H+\sum_{i}C_{i}=1$. This leads to an
explicit form of the competition $\phi $ function, namely 
\begin{equation}
\Phi (H,\mathbf{C})=G(H)+\sum_{k=1}^{M}f_{k}C_{k}
\end{equation}%
which is nothing but the average replication rate.

A numerical analysis of this system was performed for some parameter values,
showing that the population dynamics of the cancer population spread over
mutation space as a wave until a stable distribution (showing a single peak)
around high instability levels was observed. However, no systematic analysis
was performed in order to characterize potential phases and their
implications. In particular, it was not studied the behavior exhibited by
the heterogeneous population close to the optimal/lethal thresholds.
Moreover, the linear model above is an oversimplification and a better
description is needed in order to make reliable predictions.

\vspace{0.35 cm}

\section{Integral equation expansion}

The linear instability model reveals an important dynamical feature of
unstable dynamics: a propagating front is formed and moves through
instability space. Fronts (and their propagation dynamics) are a well known
characteristic of many relevant biological processes [41,44,45] and can be
analyzed in a systematic way through well known methods. Our first step here
will be to convert the discrete model presented above into a more general,
analytically tractable integral equation form. Such model will allow us
exploring the phase space of our system and to make some analytic estimates
of propagation speed.

An integral equation model can be derived starting from the previous linear
model. Let us first notice that the equations (7) for $C_{i}$ can be
re-written as 
\begin{equation}
\frac{dC_{i}}{dt}=\sum_{1}^{M}f_{j}C_{j}w_{ji}-C_{i}\Phi (H,\mathbf{C}).
\label{4}
\end{equation}%
This is done by introducing the following notation: 
\begin{equation}
w_{ji}=\delta _{j,i-1}\mu _{j}+(1-\mu _{j})\delta _{ij},  \label{3}
\end{equation}%
and, as explained in the previous section, we consider the condition $\mu_{0}=\mu_{M}=0$ to properly describe the evolution of $C_{0}$ and $C_{M}$.
An integral equation can be now constructed, using the continuous variable $%
C_{i}(t)=\Delta \mu \cdot c(\mu ,t)$. Moreover, we need to generalize the
functional connection between different instability levels, which was
assumed to be a simple function in (7) but could adopt different forms. A
general integral equation can be constructed, namely: 
\begin{widetext}
\begin{equation}
c(\mu,t+T)=c(\mu,t)+T\int\limits_{-\mu}^{0}f(\mu+\Delta _{\mu})c(\mu+\Delta_{\mu},t)\omega (\Delta _{\mu})d\Delta _{\mu}-c(\mu,t)T\phi (H,\mathbf{c}),  \label{6}
\end{equation}
\end{widetext}where we have used a continuous dispersal kernel $\omega
(\Delta _{\mu })$ [16,41,47] which provides the probability density that
cancer cells in $c(\mu -\left\vert \Delta _{\mu }\right\vert ,t)$ produce
offspring, after a given time $T$, within the $\mu $-coordinate i.e. further
cells within the $c(\mu ,t+T)$. Moreover, in Eq. (12) we have changed the notation of the average fitness from $\Phi (H,\mathbf{C})$ [as it appears in Eqs. (7) and (8)] in order to remark that we now use a continuous description for cancer cells. While in the linear model [Eqs. (7) and (8)] the average fitness depends on $H$ and $\mathbf{C}$, in our integral model [Eq. (12)] the average fitness must depend on both $H$ and the continuous distribution of cancer cells at time t (that we have writen as $\mathbf{c}$ to explicitly indicate its correspondence to $\mathbf{C}$ in the linear model). 

Following analogous steps to those for cancer cells, we can also develop the differential Eq. (8) for healthy cells so that we obtain an explicit expression for the population $H$ at time $(t+T)$. This yields:

\begin{equation}
H(t+T)=H(t) + T[G(H)-H\phi (H,\mathbf{c})]  \label{10bis}
\end{equation}%

The constant population requirement (defined above as $C+N=1$ for the mean
field model) can be expressed here as 
\begin{eqnarray}
H(t)+\int_{0}^{M}c(\mu ,t)d\mu &=&1 \\
H(t+T)+\int_{0}^{M}c(\mu ,t+T)d\mu &=&1
\end{eqnarray}%
and we assume that $M$ is large enough so that we can ensure that $c(M,t)=0$.

In this paper we will use this integral equation approach to describe our
cancer quasispecies model. This model allows us to properly study the way
the instability wave can (or cannot) propagate and some other phenomena
including the catastrophic collapse of the cancer population once the
unstable wave crosses some given thresholds.

Using the previous condition and definitions, it is possible to develop our
model equation. Let us indicate as $\phi =\phi (H,\mathbf{c})$, $f_{H}=G(H)-H\phi $ [which, considering Eq. (13), can be understood as the change in $H$ cells per unit time], and compute the total cancer cells population as:

\begin{equation}
\Lambda (t)=\int_{0}^{M}c(\mu ,t)d\mu.
\end{equation}%
Note that $\Lambda (t)$ strictly depends on $(M,t)$, but the dependence on $M$ has been omited because (as mentioned above) we consider $M$ is high enough to satisfy the condition $c(M,t)=0$. Thus, it is possible to see that our system is described by the following
mathematical expressions:

\begin{widetext}
\begin{eqnarray}
T f_{H} +H+ \Lambda (t)
+T \int\limits_{0}^{M}\int\limits_{-\mu}^{0}c(\mu+\Delta _{\mu},t)f(\mu+\Delta
_{\mu})\omega (\Delta _{\mu})d\Delta _{\mu}d\mu-T\phi \Lambda (t)= 1 \\
\Rightarrow 
1+TH\phi+T\phi \Lambda (t) =TG(H)+H+\Lambda (t)+\int\limits_{0}^{M}T\int%
\limits_{-\mu}^{0}c(\mu+\Delta _{\mu},t)f(\mu+\Delta _{\mu})\omega (\Delta_{\mu})d\Delta _{\mu}d\mu,  \label{7}
\end{eqnarray}
\end{widetext}

\bigskip From the above equation (17) it is easy to derive the following expression for the average
fitness of the population (that includes normal tissue and tumor
cells):

\begin{equation}
\phi (H,\mathbf{c})=G(H)+\int\limits_{0}^{M}\int\limits_{-\mu }^{0}c(\mu +\Delta
_{\mu },t)f(\mu +\Delta _{\mu })\omega (\Delta _{\mu })d\Delta _{\mu }d\mu .
\label{fitness}
\end{equation}

It is worth to note that the integro-difference equation (\ref{6}) permits
to analyze several dynamical properties of the system which cannot be
attained by means of the previous linear model (\ref{4}). In the linear
model, the offspring of tumor cells in a given stage $i$ may grow either in
the same stage $i$ or in the subsequent $i+1$. A desirable feature of the
continuous description from (\ref{6}) is that the dispersal kernel can
easily model different forms of instability-driven spread in the genetic
landscape. In the following section, we analyze a simple case in which
migration probability decays exponentially with the jumping distance $\Delta
_{\mu }$. The linear model can also be recovered from Eq. (\ref{6}) by
introducing a dispersal kernel that restricts mutations to discrete points
in the $\mu $-space. In order to derive some analytical solutions of the
system, such simplified dispersal kernels will be shown to be specially
useful.

\begin{figure*}[tbp]
\begin{center}
\includegraphics[width=0.7 \textwidth]{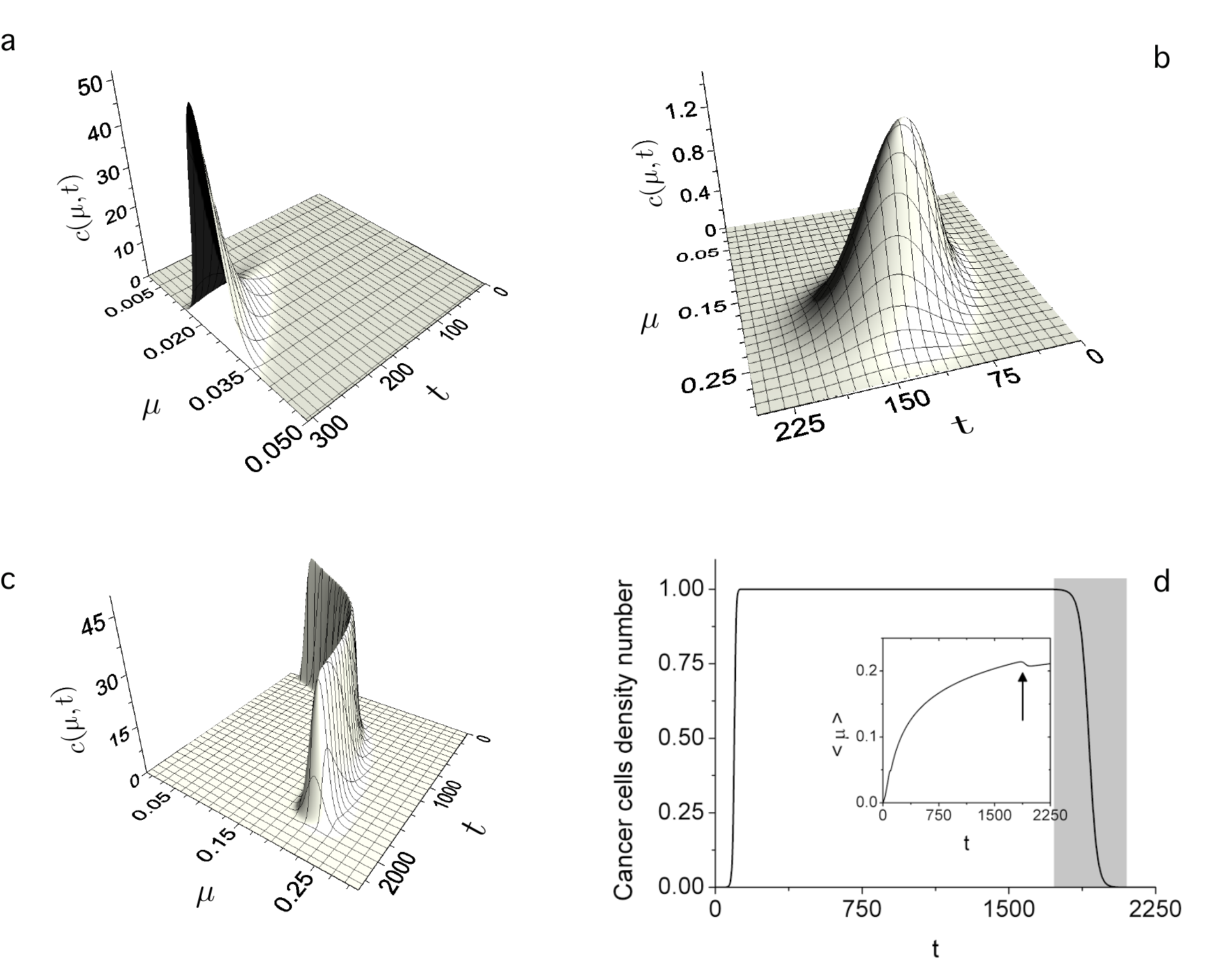}
\end{center}
\caption{The three major dynamical patterns of dynamical behaviour displayed
by our mathematical model. Here the population density for different
instability levels is plotted against instability and time. In (a), unstable
tumours expand, evolving towards a stable, high instability rate. Time
evolution for $r=0.25,\protect\alpha =20$, $\protect\mu _{c}=0.08$ and $%
\protect\mu _{disp}=3\cdot 10^{-4}$. Each time step is equivalent to a
generation of cells. Cancer cells diffuse through the instability space as a
wave. At early stages ($t<200$), the fraction of cancer cells in the system
is low. However, when cancer cells reach high enough instability (slightly
above $\protect\mu =0.015$ in this example), a rapid increase in cancer
population density is produced. (b) Tumor fails to get established. The
following parameter values have been used: $r=0.25$, $\protect\alpha =50$, $%
\protect\mu _{c}=0.08$ and $\protect\mu _{disp}= 10^{-2}.$ (c)
Population collapse. Here expansion is followed by collapse after a long
transient, as shown in (d). Here we have used $r=0.25$, $\protect\alpha =50$%
, $\protect\mu _{c}=0.08$ and $\protect\mu _{disp}=1\cdot 10^{-3}$.}
\label{Twins}
\end{figure*}

\section{Wave fronts in instability space}

In this section, we present several scenarios in which a tumor can either
collapse or succeed over a healthy tissue. According to the integral model
[Eqs. (\ref{6}) and (\ref{fitness})], tumor evolution is mainly governed by
competition. As explained above, this competition involves not only the
fight between cancer cells and healthy cells, but also the struggle within
cancer cell clones.

In the previous section we have presented a model that is mainly based on
two dynamical features of tumors: replication (introduced by the growth
function $f(\mu )$) and mutation (given by the dispersal kernel $\omega
(\Delta _{\mu })$). Concerning the replication process, below we consider
some specific growth functions involving a constant reproduction rate for
healthy cells, so that $G(H)=r_{n}H$. For tumor cells, the growth function
depends on instability as $f(\mu )=r_{n}\left( 1+\alpha \mu \right) exp(-\mu
/\mu _c)$. This was derived in [23] from the probabilistic condition defined
by equation (5). The rate $\alpha $ introduces a selective advantage for
cancer cells over healthy cells. The constant $\mu _{c}$ refers to a
characteristic instability rate.

In order to model the mutant trend of cancer cells, let us consider the
following continuous function for the dispersal kernel:

\begin{equation}
\omega (\Delta _{\mu })=\frac{1}{\mu _{disp}} \exp\left( \frac{-\left\vert\Delta _{\mu
}\right\vert}{\mu _{disp}}\right) .  \label{kernel}
\end{equation}

According to Eq. (\ref{kernel}), a parent cell generates offspring at
similar instability domains (i.e. situated at $\Delta _{\mu }\rightarrow 0$)
with higher probability than new cells presenting much higher instability
(i.e., living at $\Delta _{\mu }>>0$). The parameter $\mu _{disp}$
represents a characteristic (within generation) instability increment. Since
we have 
\[
\int_{-\infty}^{0}\omega (\Delta _{\mu })d\Delta _{\mu }=1
\]
the dispersal kernel distributes the cells of the new generation in the
instability space, but it does not modify the total number of cancer cells
in the system.

\subsection{Tumor wins phase}

Figure 3a shows the evolution of a population of cancer cells which
initially composes the $0.001\%$ of the cells in the system. Cancer cells at 
$t=0$ have been equally distributed within a range of low instability
(namely, $\mu \in (0,2\cdot 10^{-4}]$). We observe an early stage ($t\in
\lbrack 0,150]$) in which tumor cells remain at low values of the population
density $c(\mu ,t)$. Within this initial period, cancer cells do not
overcome healthy cells because their selective advantage is not significant
(i.e., $f(\mu )\simeq G(H)$ because $\mu \simeq 0$).

The dispersal kernel $\omega (\Delta _{\mu })$ pushes forward the tumor
population towards higher instability domains. In other words, at each time
step a fraction of the cancer cells offspring becomes sensibly more unstable
than their parent cells. A rapid increase in cancer cells population density
is observed about $t=200$ generations. The rapid growth affects cells whose
genetic instability is above a certain threshold (see the region above $\mu
=1.5\cdot 10^{-2}$). This indicates that such degree of instability provides
for significant selective advantage over other cells in the system. During
the fast growth phase, the population not only attains a large fraction of
the total population, but it also continues migrating (see the left to right
dispersion of the population wave). At the end of the time series in Fig.
3a, the concentration of cancer cells in the system is about $50\%$ (we
consider this condition is enough to cause the death of the host). This is
an example of the dynamics at the cancer expansion phase.

\subsection{Tumor failure phase}

It seems reasonable to think that increasing the characteristic migration
distance $\mu _{disp}$ should accelerate tumor proliferation, because cancer
cells will reach optimal instability domains faster. However, increasing $%
\mu _{disp}$ does not necessarily lead to the tumor-win phase. It can
actually jeopardize cancer propagation even when an already established
population is formed. If a tumor cell produces highly mutant descendants
(i.e., new cells accumulating many new mutations) with high probability, it
follows that the probability of generating descendants without additional
mutations cannot be very large.

Figure 3b depicts an example of the tumor-failure phase. In this case the
selective advantage presents a higher value (namely, $\alpha =50$) than that
for the tumor in the previous scenario. Here we observe a tumor population
wave diffusing in the instability space, always coexisting with normal
cells (the total number of cancer cells $\Lambda(t)$ do not exceed the $12\%$ at any generation). Despite the relatively high selective advantage $\alpha$, the high value of $\mu _{disp}$ prevents the tumor population to remain at the optimal instability domain, and hence cancer cells
cannot grow fast. The tumor moves towards excessive instability, and cancer
replication becomes smaller than that of the host tissue. These conditions
define the tumor extinction phase.

\subsection{Catastrophic tumor decay}

A qualitatively different and somewhat unexpected outcome is displayed in
Fig. 3c, where we have set a lower value of $\mu _{disp}$. As a result, a
fast extinction of healthy cells occurs and cancer cells invade all the
available space before $t=200$. Here we let the system evolve beyond the
absence of healthy cells. Even if this situation typically involves the
elimination of host cells, it could be observed in cell culture conditions.
Moreover, we need to consider a potentially relevant situation, namely when
a given tumor has expanded within large parts of the organ, as it occurs
with many malignant cancers. After the rapid increase in cancer cells
population density ($t\simeq 200$), the tumor continues its migration
towards higher instability.

Since the value of $\mu _{disp}$ is relatively high, the tumor population is
unable to stay within the optimal region. At every new generation, a large
fraction of the progeny accumulates new mutations. The final outcome is very
interesting: a collapse finally occurs. This is illustrated in figure 3d,
where we plot the total cancer population and the average instability
(inset) for the example of figure 3c. Around $1700$ generations, cancer
cells have accumulated so many mutations that they are almost unable to
produce viable descendants. After $t=2000$ there is no significant cancer
cells population.

Despite the slow growth of $\langle \mu \rangle $, a catastrophic shift
occurs, with a rapid decay of the tumor. Catastrophic shifts have been
previously described within ecological and social systems [48] and are
characterized by sudden system responses triggered by slow, continuous
changes of given external control parameters. The novelty of our observation
is that the changing parameter is affected by (and affects) population
dynamics and thus is not externally tuned but internally increased.

\begin{figure}[tbh]
\begin{center}
\includegraphics[width=0.47 \textwidth]{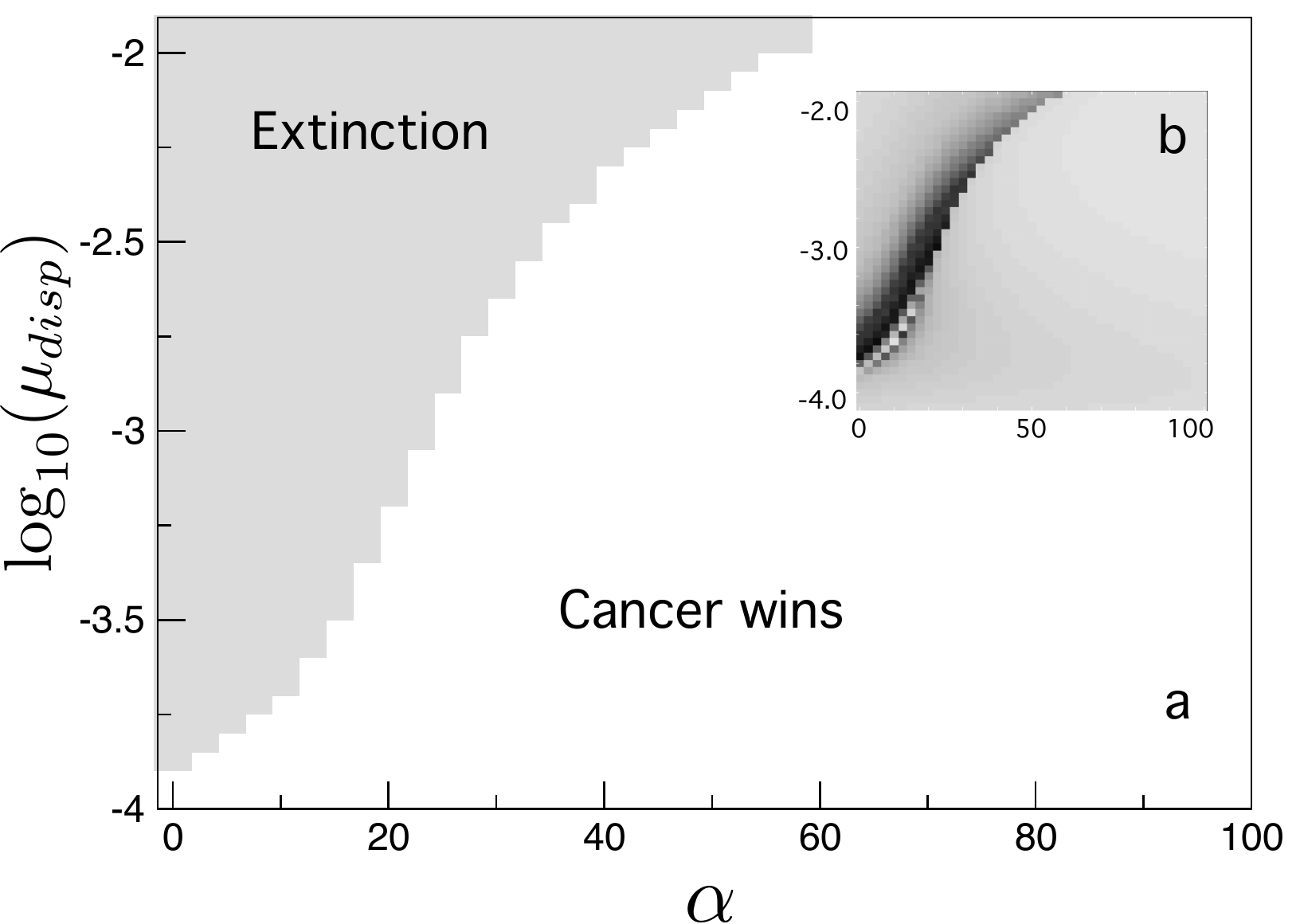}
\end{center}
\caption{Phases in the tumor growth model. The main plot (a) shows the two
phases associated to the propagation (white) or extinction (gray) of the
cancer cell population. The transition separating the two phases can be
characterized by the transient dynamics exhibited by the model. The inset
(b) displays the number of time steps (or cancer cells generations) to reach
the corresponding final state represented in \textbf{a)}. Darker (lighter)
zones are associated to longer (shorter) transients. As expected from a
phase transition phenomenon, long transients are observed close to the
boundary between both phases.}
\label{7}
\end{figure}

\subsection{Phase space}

A systematic exploration of the parameter space provides a picture of the
two main phases, as shown in Fig. \ref{7}. The two axes involve a wide range
of values for both $\alpha $ and $\mu _{disp}$. In the first phase (gray
squares), the tumor is driven to extinction. Extinction arises as a
combination of two components: \textit{i)} an insufficient fitness advantage
of the early cancer cells (the cancer population progressively decays
without reaching enough instability to develop), or \textit{ii)} the tumor
inability to keep the optimal instability (when this happens, a moderate
population growth precedes the tumor failure). The second region (white
area) stands for tumors that grow enough to overcome the healthy tissue.

The transition between the two regions is also marked by a rapid increase in
the transient time. In Fig. \ref{7}b we have depicted the transient time
steps (i.e., generations of cancer cells) to reach either the tumor
extinction or its stable expansion to equilibrium values. As expected,
longer times are needed near the phase transition.

\section{Tumor front speed}

In the previous section, we have seen how some tumor population waves
diffuse in the instability space. A relevant feature of propagating fronts,
with direct importance for tumor growth, is the propagation speed of the
front. Such speed has been actually calculated for spatially growing tumors
[17,49,50] and the front is thus a spatially defined one. Although we are
here considering front propagation through instability space, the same
reasoning applies. Here we derive an analytical, approximate solution for
the front speed of the tumor. This will provide a quantitative measure of
how fast cancer instability propagates. Since deriving an exact analytical
expression for the front speed can be extremely cumbersome, some
approximations are required.

First, let us consider early stages in tumor development (such as the first $%
150$ in Fig. \ref{Twins}). Here the system is mostly composed of healthy
cells, and few of cancer cells. This permits to approximate the complex
expression for the average fitness [see Eq. (\ref{fitness})] as the
reproduction rate of healthy cells, i.e.,%
\begin{equation}
\phi (H,\mathbf{c})\simeq G(H)\simeq r_{n}.  \label{early}
\end{equation}%
The second approximation we will consider refers to the dispersal kernel.
According to Eq. (\ref{kernel}) in the previous section, the dispersal
kernel is a continuous function defined in the interval $[-\infty,0]$. In
this section we will consider the following simpler, discrete dispersal
kernel:

\begin{equation}
   \omega (\Delta _{\mu })=2p_{e}\delta (\Delta _{\mu })+(1-p_{e})\delta (\Delta_{\mu }+\mu _{disp})\gamma(\mu_{disp}),  \label{dirackernel}
\end{equation}

where $\delta (\Delta _{\mu })$ corresponds to the Dirac delta function
operating on the variable $\Delta _{\mu }$, and $\gamma(\mu_{disp})=1$ if $\mu _{disp}\cancel{=} \mu$ and $2$ otherwise [51].

The above discrete kernel (\ref{dirackernel})
considers that every new cell can either stay at the same instability $\mu $
of the parent cell (with probability $p_{e}$, which is called persistence)
or jump into a higher instability $\mu +\mu _{disp}$ [with probability $%
(1-p_{e})$]. Although the discrete kernel (\ref{dirackernel}) is much
simpler than the continuous kernel (\ref{kernel}), it also models a major
feature in cancer cells replication (see the previous section), that is: the
stronger the mutant trend of cancer cells, the weaker the ability of the
population to keep an optimal instability.

Thus, according to Eqs. (\ref{early}) and (\ref{dirackernel}) above, our
approximation to Eq. (\ref{6}) reads:

\begin{widetext}
\begin{equation}
c(\mu,t+1)=c(\mu,t)+\int\limits_{-\mu}^{0}f(\mu+\Delta _{\mu})c(\mu+\Delta_{\mu},t)(2p_{e}\delta(\Delta_{\mu})\\
+(1-p_{e})\delta(\Delta_{\mu}+\mu_{disp})\gamma(\mu_{disp}))d\Delta _{\mu}-c(\mu,t)r_{n}.  \label{dirac1}
\end{equation}
\end{widetext}

Taking into account the integrative properties of the Dirac delta function $%
\delta (\Delta _{\mu })$, the above equation \ref{dirac1} can be rewritten
in terms of a much simpler functional form: 
\begin{widetext}
\begin{equation}
\begin{array}{c}
c(\mu ,t+1)=c(\mu ,t)+p_{e}c(\mu ,t)f(\mu ) 
+(1-p_{e})c(\mu -\mu _{disp},t)f(\mu -\mu _{disp})-c(\mu ,t)r_{n},%
\end{array}
\label{diracmodel}
\end{equation}
\end{widetext} where we have assumed that the condition $\mu\geq\mu_{disp}$ holds, since we are interested in the propagation of the tumor front. Indeed, for $\mu<\mu_{disp}$ cancer cells evolve as in Eq. (\ref{diracmodel}) but neglecting the third term on the RHS. The front speed from reaction-dispersal integro-difference
equations such as (\ref{6}) can be obtained under some general assumptions
[16,17] associated with the shape to be expected for the propagating front.

Here we are interested in the simplified version (\ref{diracmodel}) of the
model. Thus we only need to assume that there exist constant shape solutions
of the form 
\[
c(\mu ,t)=c_{0}\exp \left [ -\lambda z \right] 
\]
for large values of the coordinate $z \equiv (\mu -vt)$.

This yields the following approximate,
analytic relation between the tumor front speed and the wave front shape parameter $\lambda$:  
\begin{widetext}
\begin{equation}
v(\lambda)= \frac{1}{\lambda }\ln \left [ p_{e}f(\mu ) 
+(1-p_{e})f(\mu -\mu _{disp}) e^{\lambda \mu _{disp}} -r_{n}+1 \right ] .
\label{speed}
\end{equation}
\end{widetext}
\noindent Finally, by means of the standard, marginal stability condition [16] the minimal speed $v_{*}=\min_{\lambda >0}v(\lambda)$ is the one selected by the front. In the analysis below, the following method has been used to find the value of the approximate speed $v_{*}$. First, we plotted the numerical solution to Eq. (25) on the $\lambda >0$ axis. As usual, we obtain a function $v(\lambda)$ that is convex from below [16]. Thus, we look for the minimum value of this function, which reveals the approximate value for the front speed of the tumor as it propagates through instability space (i.e., $v_{*}$).

\begin{figure}[tbh]
\begin{center}
\includegraphics[width=0.5 \textwidth]{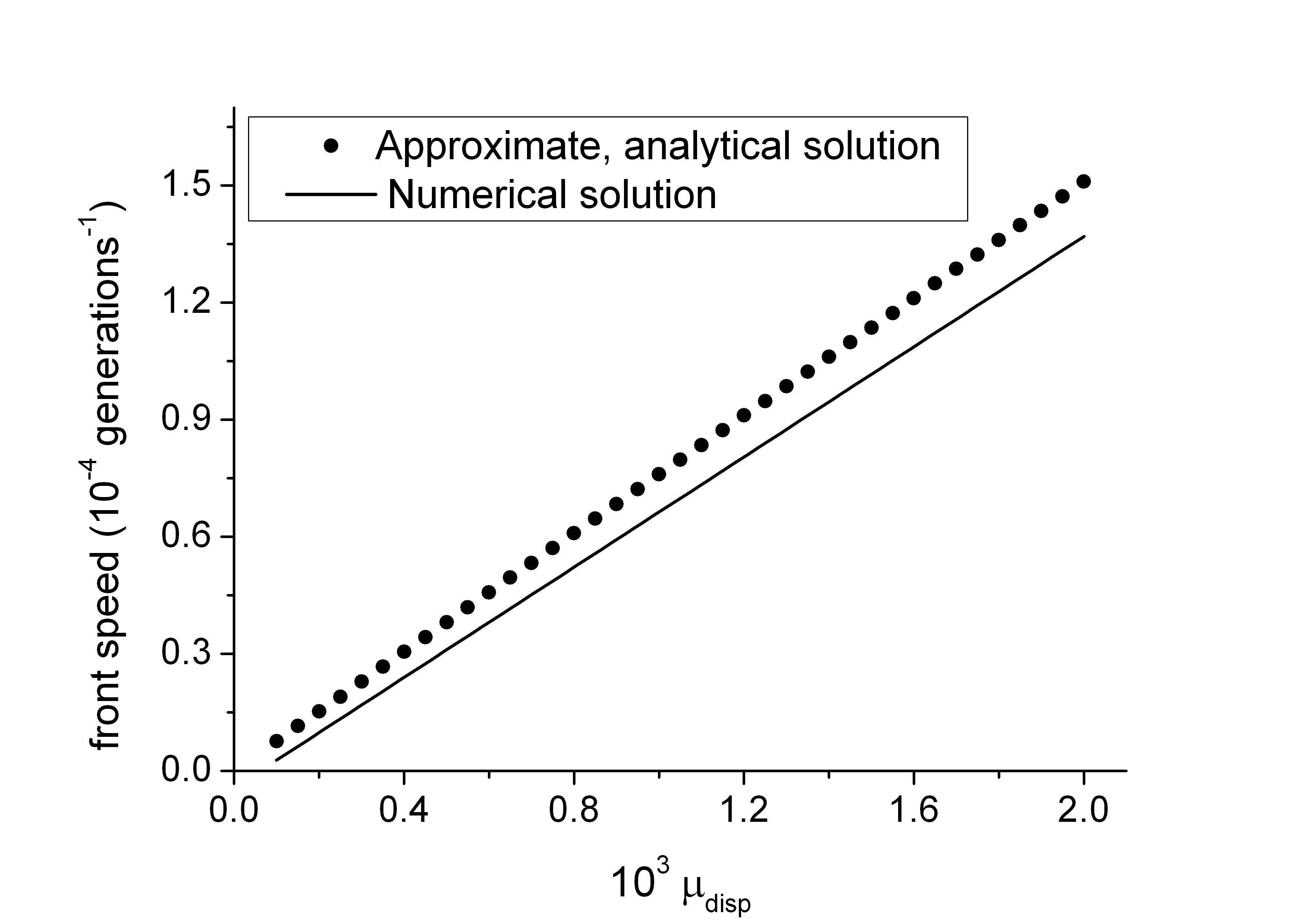}
\end{center}
\caption{Front speed of the cancer population travelling on the instability
space, as a function of the characteristic dispersal distance $\protect\mu %
_{disp}$. The line and the circles stand for the numerical results and the
approximate analytical speed $v_{*}$, respectively. The rest of the parameters used to compute the front
speed are: $r_{n}=0.25$, $\protect\alpha =20$, $p_{e}=0.85$ and $\protect\mu %
_{c}=0.08.$}
\label{8}
\end{figure}

The approximate front speed $v_{*}$ should not be taken as a general
trend in tumor evolution, since it is subject to the approximations
explained above. Indeed, for cases in which healthy cells overcome the tumor
it eventually predicts negative values of the front speed. However,
predicting a negative front speed can also be seen as the retreat (i.e., the
death) of the cancer population (which at early times is only composed by a
few cancer cells with $\mu \rightarrow 0$). Nevertheless, the approximate speed $v_{*}$
provides remarkably good results for the front speeds of lethal tumors
(i.e., for tumors within the parameter region in which the tumor succeeds),
as we show in Fig. \ref{8}.

Fig. \ref{8} shows a comparison between the numerical and the approximate
analytical speed $v_{*}$ for the tumor front speed as a
function of the characteristic dispersal distance $\mu _{disp}$. Numerical
solutions for the front speed have been computed by numerically solving [52]
the model Eqs. (\ref{6}) and (\ref{fitness}) using the discrete version of
the dispersal kernel (\ref{dirackernel}). For both the numerical and the
approximate analytical solutions, the front speed monotonically increases
with the characteristic distance $\mu _{disp}$. As far as the order of
magnitude is concerned, the approximate analytical speed $v_{*}$
is able to predict the more exact numerical results for the tumor front
speed. Furthermore, relative differences (which are typically above $15\%$)
between the analytical results and the numerical solutions are approximately
independent of $\mu _{disp}$.

The solutions for the front speed on the instability space in Fig. \ref{8}
exhibit an approximately linear dependence on the characteristic dispersal
distance $\mu _{disp}$. Since the instability $\mu $ of tumor cells is
directly related with the selection of cells during tumor evolution, $\mu
_{disp}$ can also be interpreted as a parameter that determines the
selective intensity on cancer cells. Previous models on adaptive front waves
have also yielded adaptation speeds that depend on selective intensity
parameters. In this context, in [14] the speed of adaptation for asexual
clones that compete for space scales as $\mu ^{1/2}$ in one-dimensional
habitats and $\mu ^{1/3} $ in two-dimensional ones (in their model, $\mu $
stands for the rate at which new mutations appear at each lattice site).
Also, in Ref. [53] the authors found that, when a continuous two-dimensional
space is considered, genetic wave speeds are proportional to $s^{1/2}$,
where $s$ represents the small fitness effects $s$ of beneficial mutations.
In our model we have studied the front speed of the tumor on the instability
space, without taking into account neither spatial structure nor the
physical environment. However, both $\mu $ and $\mu _{disp}$ have similar
interpretations to the selective parameters in Refs. [14,53]. Future work
could be directed to introduce spatial effects into our model, and it would
be interesting to explore if similar scaling exponents arise in the
dependence of the tumor invasion speed and genetic instability rates.

\section{Discussion}

In this paper we have presented an integral model for the evolution of
unstable tumors. Our model improves a previous compartment description of
the cancer cells population, because we consider the genetic instability as
a continuous variable that characterizes the state of the cell. The model
considers a population of tumor cells that replicate and migrate (mutate) in
the instability space, while competing for available resources (a limited
population constraint has been applied). This model is based on several
simplifying assumptions, from the linear nature of interactions between
instability levels to the dispersal kernels used.

We have presented an extended analysis of unstable cancer evolution over the
two most relevant parameters of the model: the selective advantage of cancer
cells over the healthy cells population, and the characteristic migration
distance within instability space (which determines the mutant tendency of
cancer cells). Several outcomes of the process have been found. Two of them
are expected: either the growth or the failure of cancer to succeed are
predicted by the simplest mean field model that can be defined, as discussed
in section II. The integral equation approach confirms such prediction,
although it allows to substantiate it in more accurate ways, providing a
formal framework to calculate useful quantities, particularly the front
speed of our population through the $\mu $-space. Moreover, this formal
approach provides a natural way to properly introduce population
heterogeneity.

An additional scenario has also been found, namely the catastrophic shift
phase, where the tumor grows, eventually expanding over a significant part
of the total available space, with a steady growth of instability. However,
at some point the excessive instability level leads to a population
collapse, with no cancer cells in the end. Our model does not consider immune components or other biologically relevant
factors [54]. Instead, the key factor responsible for the tumor collapse is high
genetic instability. This result provides further support to the original
proposal that lethal thresholds of instability exist in cancer [23,24] which
could be exploited for therapeutic purposes, even when major success of the
cancer population is observable. Future work should further explore this
observation, adding also other known threats to cancer progression, such as
starvation or hypoxia, which could further enhance the frequency and
sharpness of these thresholds.


\begin{acknowledgments}

We thank the members of the CSL for useful discussions. This work was
supported by grants from the Fundacion Botin, a MINECO grant and the Santa
Fe Institute.
\end{acknowledgments}

-----------------------------------------------------------------------------

\bigskip

\section{References}

\begin{enumerate}
\item M. Greaves and C. C. Maley, Nature \textbf{481}, 306 (2012).

\item L. M. F. Merlo, J. W. Pepper, B. J. Reid and C. C. Maley, Nature Rev
Cancer \textbf{6}, 924 (2006).

\item B. Vogelstein and K. W. Kinzler, Nature Medicine \textbf{10}, 789
(2004).

\item D. P. Cahill, K. W. Kinzler, B. Vogelstein and C. Lengauer, Trends
Genet. \textbf{15}, M57 (1999).

\item M. Chow and H. Rubin, Cancer Res. \textbf{60}, 6510 (2000).

\item J. Jackson L. A. Loeb Genetics \textbf{148}, 1483 (1998).

\item C. Lengauer, K. W. Kinzler and B. Vogelstei, Nature \textbf{396}, 643
(1998).

\item L. A. Loeb, Cancer Res. \textbf{51}, 3075 (1991).

\item L. A. Loeb, Cancer Res. \textbf{54}, 5059 (1994).

\item B. Vogelstein, N. Papadopoulos, V. E. Velculescu, S. Zhou, L. A. Diaz
Jr. and K. W. Kinzler, Science \textbf{339}, 1546 (2013).

\item F. Michor, Y. Iwasa and M. A. Nowak, Nat. Rev. Cancer \textbf{4}, 197
(2004).

\item N. Beerenwinkel, T. Antal, D. Dingli, A. Traulsen, K. Kinzler, V. E.
Velvulescu, B. Vogelstein and M. A. Nowak, PLOS Comp. Biol. \textbf{3}, e225
(2007).

\item R. V. Sol\'{e} and J. Bascompte, \emph{Self-Organization in Complex
Ecosystems} (Princeton U. Press, Princeton, New Jersey, 2007).

\item E.A. Martens and O. Hallatschek, Genetics \textbf{189}, 1045 (2011).

\item E.A. Martens, R. Kostadinov, C. C. Maley and O. Hallatschek, New J.
Phys. \textbf{13}, 115014 (2011).

\item J. Fort and T. Pujol, Rep. Prog. Phys. \textbf{71}, 086001 (2008).

\item J. Fort and R. V. Sol\'{e}, New J. Phys. \textbf{15}, 055001 (2013).

\item E. Khain, L. M. Sander, and A. M. Stein, Complexity \textbf{11}, 53
(2005).

\item R. A. Gatenby and E. T. Gawlinski, Cancer Res. \textbf{56}, 5745
(1996).

\item S. Negrini, V. G. Gorgoulis and T. D. Halazonetis, Nat. Rev. Mol. Cell
Biol. \textbf{11}, 220 (2010).

\item R. A. Gatenby and B. R. Frieden, Cancer Res. \textbf{62}, 3675 (2002).

\item R. A. Gatenby and B. R. Frieden, Mutat. Res. \textbf{568, }259 (2004).

\item R. V. Sol\'{e}, Europ. J. Phys. B \textbf{35}, 117 (2003).

\item R. V. Sol\'{e} and T. Deisboeck, J. Theor. Biol. \textbf{178}, 47
(2004).

\item J. M. Coffin, Science \textbf{267}, 483 (1995).

\item E. Domingo, J. J. Holland, C. Biebricher and M. Eigen, in: \emph{%
Molecular Evolution of the Viruses}, edited by A. Gibbs, C. Calisher and F.
Garcia-Arenal (Cambridge U. Press, Cambridge, 1995).

\item Domingo, E. and Holland, J. J. in: \emph{The evolutionary biology of
RNA viruses}, pp. 161-183, edited by S. Morse (Raven Press, New York, 1994).

\item M. Eigen, Naturwiss. \textbf{58}, 465 (1971).

\item M. Eigen, J. McCaskill and P. Schuster, Adv. Chem. Phys. \textbf{75},
149 (1987).

\item P. Schuster, in: \emph{Complexity: metaphors, models and reality}, pp.
383-418, edited by G. A. Cowan, D. Pines and D. Meltzer (Addison-Wesley,
Reading, MA 1994).

\item S. Cottry, C. E. Cameron, and R. Andino, Proc. Natl. Acad. Sci. USA 
\textbf{98}, 6895 (2001).

\item L. A. Loeb, J. M. Essigmann, F. Kazazi, J. Zhang, K. D. Rose and J. I.
Mullins, Proc. Natl. Acad. Sci. USA \textbf{96}, 1492 (1999).

\item J. J. Holland, E. Domingo, J. C. de la Torre and D. A. Steinhauer, J.
Virol. \textbf{64}, 3960 (1999).

\item P. R. A. Campos and J. F. Fontanari, PRE \textbf{58}, 2664 (1998).

\item P. Tarazona, Phys. Rev. A \textbf{45}, 6038 (1992).

\item P. J. Gerrish, A. Colato, A. S. Perelson , and P. D. Sniegowski, PNAS 
\textbf{104}, 6266 (2007).

\item L. A. Loeb, Nat. Rev. Cancer \textbf{11}, 450 (2009).

\item Q. Zhang and R. H. Austin, Annu. Rev. Cond. Mat. Phys \textbf{3}, 363
(2012).

\item R. Pastor-Satorras and R. V. Sol\'{e}, Phys. Rev. E \textbf{64},
051909 (2001).

\item D. B. Saakian, E. Munoz, C. K. Hu, and M. W. Deem. Phys. Rev. E 
\textbf{73}, 041913 (2006).

\item V. Ortega-Cejas, J. Fort and V. Mendez, Ecology \textbf{85}, 258
(2004).

\item E. Eisenberg and E. Y. Levanon, Trends Genet. \textbf{19}, 362 (2003).

\item P. A. Futreal, L. Coin, M. Marshall, T. Down, T. Hubbard, Richard
Wooster, N. Rahman and M. R. Stratton, Nature Rev. Cancer \textbf{4}, 177
(2004).

\item I. Gonzalez-Garcia, R. V. Sol\'{e} and J. Costa, Proc. Natl. Acad.
Sci. USA \textbf{99}, 13085 (2001).

\item A. Marusyk, V. Almendro, K. Polyak, Nat. Rev. Cancer \textbf{12}, 323
(2012).

\item R. V. Sol\'{e}, C. Rodriguez-Caso, T. Deisboeck, and J. Saldanya, J.
Theor. Biol. \textbf{253}, 629 (2008).

\item N. Isern, J. Fort and J. P\'{e}rez-Losada, J. Stat Mech: Theor Exp 
\textbf{10}, P10012 (2008).

\item M. Scheffer, \emph{Critical transitions in Nature and Society}
(Princeton U Press, New York 2009); R. Sol\'{e}, \emph{Phase Transitions}
(Princeton U Press, New York, 2011).

\item K. R. Swanson, C. Bridge, J. D. Murray, and J. E. C. Alvord, J.
Neurol. Sci. \textbf{216}, 1 (2003).

\item K. R. Swanson, J. E. C. Alvord, and J. D. Murray, Math. Comp. Modell. 
\textbf{37}, 1177 (2003).

\item Note that the $2$ factors in Eq. (22) provide the necessary corrections in case the arguments of the Dirac delta functions in Eq. (22) coincide with the integration limits in Eqs. (12) or (24). On the one hand, the factor $2$ appearing explicitly in Eq. (22) is necessary for the dispersal kernel to satisfy the condition $\int_{-\infty}^{0}\omega (\Delta _{\mu })d\Delta _{\mu }=1$ (see Section V). For an analogous reason we consider the factor $2$ in $\gamma(\mu_{disp})$, so that cells at $c(\mu_{disp},t+1)$ receive the proper fraction [i.e., $(1-p_{e})$ instead of $(1-p_{e})/2$] from the parent cells at $c(0,t)$.   

\item When numerically solving the integral model, the position of the edge
of the front can be obtained for each generation (time step). The numerical
solutions for the front speed (Fig. 5) have been obtained by linear
regression of the time-dependent position of the front. For proper
comparison, all the analytical solutions and the numerical results in Fig. 5
have been computed at transient times such that the edge of the front is
close to $\mu =0.01$.

\item E. S. Claudino, M. L. Lyra and I. Gleria, Phys. Rev E \textbf{87},
032711 (2013).

\item R. P. Garay and R. Lefever, J. Theor. Biol. \textbf{73}, 417 (1978).

\end{enumerate}

\end{document}